\documentclass[10pt,letterpaper,twocolumn]{article} 

\pdfoutput=1

\usepackage{ol2}
\usepackage{amsmath}
\usepackage{setspace}
\usepackage{enumitem}

\begin{document}

\twocolumn[ 

\title{Stable Optical Phase Modulation with Micromirrors}

\author{Caleb Knoernschild, Taehyun Kim, Peter Maunz, Stephen Crain, Jungsang Kim$^*$}
\address{
Fitzpatrick Institute for Photonics, Electrical and Computer Engineering Department, Duke University, Durham NC, 27708\\
$^*$Corresponding author: jungsang@duke.edu
}


\begin{abstract*} 
We measure the motional fluctuations of a micromechanical mirror using a Michelson interferometer, and demonstrate its interferometric stability. The position stability of the micromirror is dominated by the thermal mechanical noise of the structure. With this level of stability, we utilize the micromirror to realize an ideal optical phase modulator by simply reflecting light off the mirror and modulating its position. The resonant frequency of the modulator can be tuned by applying a voltage between the mirror and an underlying electrode. Full modulation depth of $\pm \pi$ is achieved when the mirror resonantly excited with a sinusoidal voltage at an amplitude of 11V.

\end{abstract*}


] 

\begin{figure}[b]
	\centering
	\includegraphics[width=3in]{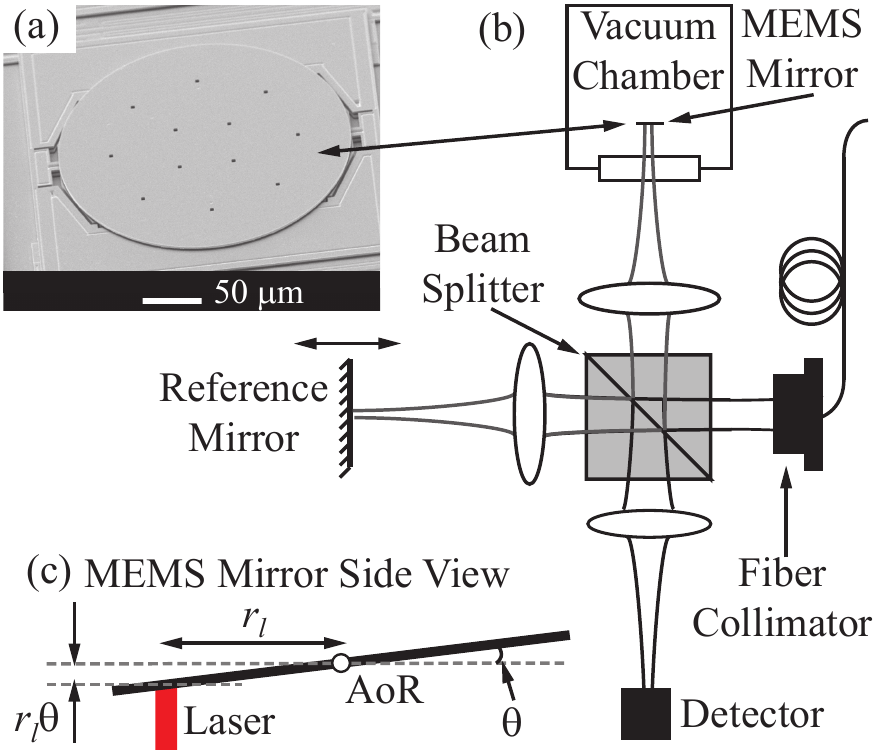}
	\caption[MEMS mirror stability measurement system schematic]{Schematic of the measurement setup. (a) Scanning electron micrograph of a MEMS mirror. (b) MEMS mirror stability measurement setup using a Michelson interferometer. (c) Shifting the reflection point on the MEMS mirror for sensitivity in the tilting mode. }
	\label{fig:PhaseSchematic}
\end{figure}

Micromirrors fabricated using micro-electromechanical systems (MEMS) technology are well suited to steer a potentially large number of laser beams. MEMS systems are used in all-optical switching \cite{NeilsonJoLT2004,KimIPTL2003}, scanning projection displays \cite{ConantSAAA2000,Castelino2005}, and optical addressing of quantum information processing systems \cite{KnoernschildOE2009,KnoernschildAPL2010}. Compared to traditional beam steering techniques such as acousto-optic (AO) and electro-optic (EO) deflectors \cite{GottliebBook1983,SchmidtKalerAPB2003,KimAO2008}, MEMS-based approaches typically consume much less actuation energy and device volume per beam. These advantages make scalable systems capable of simultaneously steering a large number of beams possible. In many applications in quantum information processing with atoms, the optical phases of the multiple beams addressing the atomic qubits have to be maintained \cite{WangPRA2010}. In this paper, we present a characterization of the pointing and phase stability of MEMS mirrors and show that carefully designed mirrors can easily achieve interferometric stability. In addition, such a mirror can be used to realize an ideal optical phase modulator.

We consider a micromirror consisting of a circular polysilicon mirror plate tethered to anchors with two torsional springs [Fig. \ref{fig:PhaseSchematic}(a)] \cite{ChangsoonSTiQEIJo2007}. The MEMS structure is fabricated using Sandia's SUMMiT V MEMS foundry process, where alternating layers of polysilicon (structural and routing layers) and silicon dioxide (SiO$_2$: sacrificial layers) are used to form the devices. At the end of the fabrication process, the SiO$_2$ is selectively removed to enable free motion of the polysilicon structural layers. The released mirror is electrostatically actuated by applying a voltage between the grounded mirror plate and the underlying electrodes.
The mirror structure features several mechanical modes of vibration with the lowest resonant frequencies belonging to the tilting and sagging modes. The behavior of these modes can be captured using a damped harmonic oscillator model:
\begin{eqnarray}
I\ddot{\theta} + D_\theta \dot{\theta} + 2 \kappa \theta & = & \frac{1}{2} \frac{\partial C (\theta,z)}{\partial \theta} V^2 +F_{\theta}, \label{equ:harmonic1}\\
M\ddot{z} + D_z \dot{z} + 2Kz & = & \frac{1}{2} \frac{\partial C(\theta, z)}{\partial z} V^2 + F_{z},
\label{equ:harmonic2}
\end{eqnarray}
where $\theta$ and $z$ are the variables describing the tilt angle and vertical sag of the mirror plate, respectively. Variables $I$ ($M$), $D_\theta$ ($D_z$), $\kappa$ ($K$), $C(\theta, z$), $V$ and $F_{\theta(z)}$ denote the rotational inertia (mass) of the mirror plate, damping coefficient for the tilting (sagging) mode, torsional stiffness (spring constant) of the torsional springs, capacitance between the mirror plate and the underlying actuation electrode, applied voltage between the electrode and the mirror plate and other driving forces for the tilting (sagging) mode, respectively. The mirror features a tilting (sagging) resonance in the $\theta$ direction at $w_\theta = \sqrt{2\kappa / I}$ ($z$ direction at $w_z = \sqrt{2K/M}$). Modes with higher resonant frequencies exist that involve higher-order distortion of the mirror plate. The dominant damping mechanism for the mirror motion is squeeze film damping by the air between the mirror plate and the substrate \cite{AndrewsSaAAP1993}. For optimal beam steering performance, the mirror is designed to feature near critical damping for the tilt motion at atmospheric pressures \cite{ChangsoonSTiQEIJo2007,KnoernschildOL2008}. The damping can be reduced by either placing the mirror in vacuum, or by modifying the device design to provide smaller mirror radius and larger gap between the mirror and the substrate.

We measured the pointing stability of this MEMS mirror in a Michelson interferometer, as shown in Fig. \ref{fig:PhaseSchematic}(b). A single mode, $\lambda = 780$ nm wavelength laser is directed to a $50/50$ beam splitter. Half of the light is sent toward the MEMS device in the sample arm of the interferometer and the other half onto a standard mirror in the reference arm. The light in the sample arm is focused by a $50$ mm focal length lens to a beam waist of $15$ $\mu$m at the MEMS mirror. The MEMS mirror has a radius of $R=130$ $\mu$m and resonant frequency of $w_\theta /2 \pi =140$ kHz for the tiling mode ($w_z/2 \pi = 258$ kHz for sagging mode). It is housed inside a vacuum chamber to control the damping. Light in the reference arm of the interferometer travels through a second $50$ mm focal length lens before reflecting off the standard flat mirror. The standard mirror is mounted on a kinematic mount with a piezoelectric actuator to tune the optical path length difference of the interferometer. Light from the two arms recombines on the beam splitter before a final lens focuses it onto a photodetector. The optical power at the detector, assuming no losses, follows the standard interference equation where detected current $I_{det} = I_{opt} [1+ \cos (2 \pi d_{opt} / \lambda + \phi_i) ] /2$, where $I_{opt}$ is the optical power used in the interferometer, $d_{opt}$ is the optical path length introduced by the micromirror motion, and $\phi_i$ is the bias point of the interferometer determined by the position of the standard mirror \cite{Saleh}. The interferometer is actively stabilized at $\phi_i = \pi/2$ to maximize sensitivity.

The sagging mode of the mirror changes the path length in the sample arm by slightly shifting the reflection plane along the direction of propagation. The tilting motion induces a tilt in the sample arm beam path, and does not lead to an optical path length change in the interferometer. A path length change using the tilting mode is realized by aligning the beam to reflect off the mirror a distance $r_l \approx 0.75R$ away from the axis of rotation [Fig. \ref{fig:PhaseSchematic}(c)], leading to $d_{opt}=2 r_l \theta$.

\begin{figure}[tb]
	\centering
	\includegraphics[width=3.25in]{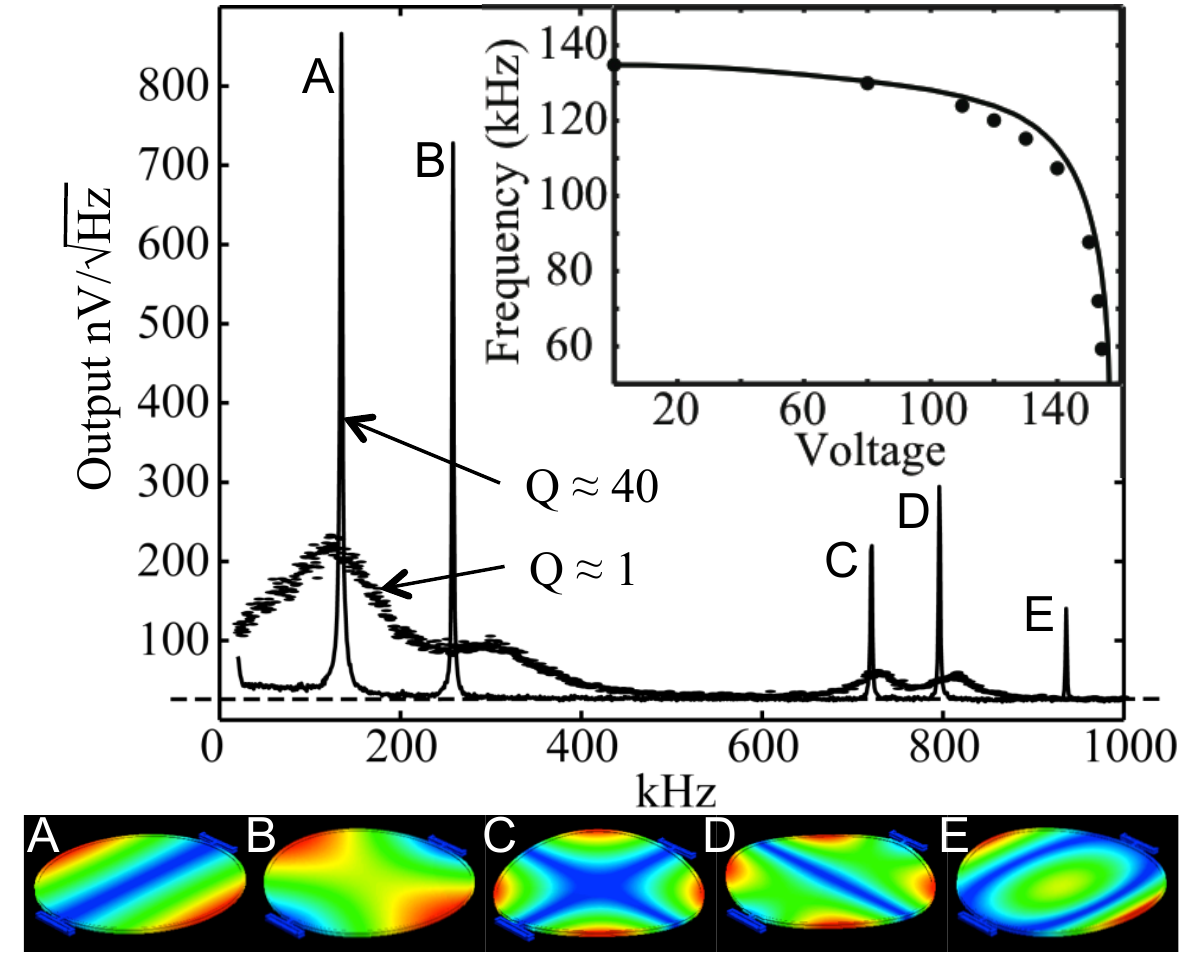}
	\caption[Thermal noise spectrum of MEMS mirror with increasing DC Voltage]{Interferometer output representing the position noise spectrum of a micromirror at a pressure of $5$ Torr ($Q\approx40$) and $760$ Torr ($Q\approx 1$) with 0V bias voltage. The dotted line shows shot noise level corresponding to laser intensity fluctuations. Mode shape corresponding to the five peaks (A -- E) are shown by a color plot of maximum displacement. The inset shows tuning of tiling mode resonant frequency as a function of bias voltage. }
	\label{fig:PhaseStab}
\end{figure}

Figure \ref{fig:PhaseStab} shows the noise spectrum of the interferometer output, corresponding to the phase noise arising from the motion of the micromirror. The two curves correspond to operating the mirror in a background pressure of $\sim 5$ and 760 Torr with a corresponding quality factor $Q=\sqrt{2 \kappa I}/D_\theta \approx 40$ and 1 for the primary tilting mode, respectively. The photocurrent from the detector is converted to a voltage signal using a transimpedance amplifier with tranimpedance gain of $R_f$ = 2 kOhms. The detected photocurrent of $I_{ph} = 0.6$mA corresponds to a shot noise level of $\sqrt{2qI_{ph}}R_f \approx 28$nV/$\sqrt{\mathrm{Hz}}$ (dashed line). This shot noise value describes the fundamental sensitivity of the interferometer's ability to detect a phase shift with a noise level of $\sim 5\times10^{-8}$ rad/$\sqrt{\mathrm{Hz}}$.

Fluctuations in micromirror position is primarily driven by thermal mechanical noise force \cite{DjuricMR2000} with (frequency-independent) power spectral density of $\tilde{F}_{\theta(z)}^2(\omega) = 4k_B T D_{\theta(z)}$. In this expression, $\tilde{F}_{\theta(z)}$ is the Fourier transform of the driving force $F_{\theta(z)}$, $k_B$ is the Boltzmann constant, and $T$ is the operating temperature of the mirror. The position of the mirror responds to this driving force according to Eq. (\ref{equ:harmonic1}) and (\ref{equ:harmonic2}). The peaks in Fig. \ref{fig:PhaseStab} show the resulting position noise corresponding to five lowest normal modes of the micromirror (labeled A -- E, the maximum displacement of the mode shown at the bottom). When integrated over the full spectrum, thermal noise from each mechanical mode will correspond to an equivalent energy of $k_BT/2$ due to equipartition theorem. From the torsional stiffness of $2\kappa = 1.7 \times 10^{-6}$Nm (spring constant of $2K=1.4\times 10^3$N/m) extracted from the resonant frequency, the integrated RMS tilt angle (position) noise from main tilting mode is $\sim5 \times 10^{-8}$rad ($\sim 1.7$pm for sagging mode). Converting the tilt angle noise to interferometer output using the respective $Q$ values, the size of noise peaks A in Fig. \ref{fig:PhaseStab} is consistent with this estimate within a factor of 2. From this analysis, we conclude that the total intrinsic noise from the micromirror contributing to the RMS phase noise of the optical field is less than $10^{-4}$ rad.

When a DC bias voltage is applied between the mirror and underlying electrode, the resonance frequency of the mechanical oscillator shifts to a lower value. The inset in Fig. \ref{fig:PhaseStab} illustrates this effect known as electrostatic softening for the tilting mode. This mechanism can be used to tune the resonant frequency of a given mechanical oscillator.

\begin{figure}
	\centering
	\includegraphics[width=3.25in]{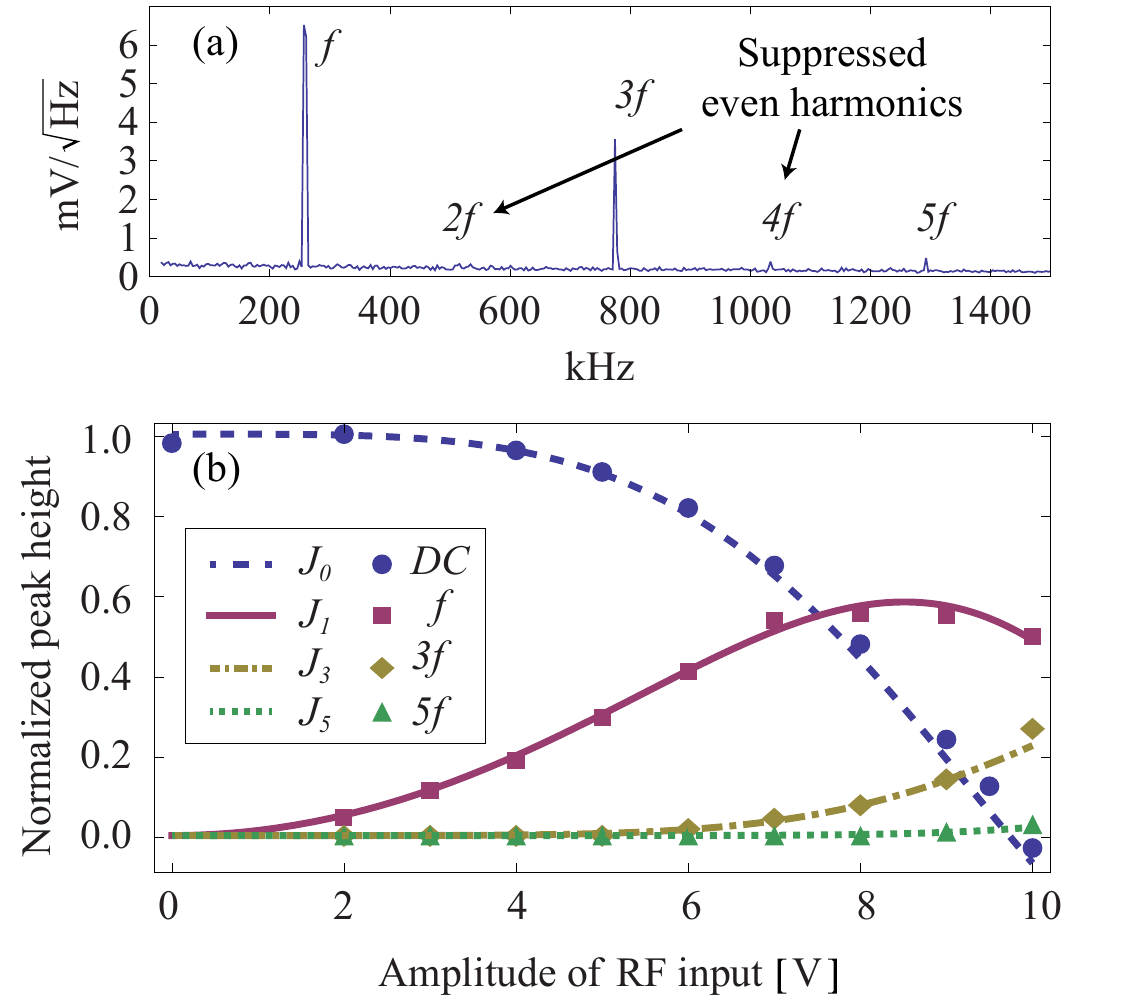}
	\caption{(a) Spectrum of interferometer output for a given modulation amplitude of a sagging mode at $f=$258 kHz. (b) Normalized spectrum peak heights as a function of RF voltage amplitude. 
					Bessel functions of different orders $J_n(s \cdot v_{amp}^2)$ are plotted using a single fitting parameter $s=0.025 / V^2$ (lines), while the measured DC values (circles) and peak heights are shown with squares ($f$), diamonds ($3f$), and triangles ($5f$).
					}
	\label{fig:BesselFit}
\end{figure}

The demonstrated stability of the mirror suggests that the resonant modes of the micromirrors can be utilized to modulate the phase of an optical field at radio frequency (RF) with low additional phase noise.
The depth of modulation for a micromirror driven at mechanical resonance is amplified by the quality factor $Q$, enabling significant modulation with only modest driving voltages. Furthermore, the resonant frequency of the micromirror can be tuned using electrostatic softening.
We characterized the optical phase modulation properties with the Michelson interferometer [Fig. \ref{fig:PhaseSchematic}(b)] using the sagging mode at 258kHz and reflecting the laser beam at the center of the micromirror.
When the optical path length difference is modulated by $(\beta \lambda / 2\pi)\sin \omega t$, the detected current $I_{det}$  is given by
\begin{align}
I_{det}&\propto 1+ \cos( \phi_i + \beta\sin \omega t) =  1+ J_0(\beta)\cos\phi_i \nonumber\\
& +  2\{J_2(\beta)\cos 2\omega t + J_4(\beta)\cos 4\omega t +\cdots \}\cos\phi_i \label{Eq:BesselExpansion}\\
& -  2\{J_1(\beta)\sin  \omega t + J_3(\beta)\sin 3\omega t +\cdots \}\sin\phi_i.\nonumber
\end{align}
In the experiment, we set $\phi_i = \pi/2$ to suppress all even order modulation sidebands. Due to quadratic dependence on the applied voltage shown in Eq. \ref{equ:harmonic2}, we use a RF voltage at frequency $f/2$ to induce a phase modulation at frequency $f$. Depth of the optical path length modulation is proportional to the square of the amplitude of the driving RF voltage $v_{d}$, so we let $\beta \equiv s \cdot v_{d}^2$ to define the proportionality constant $s$. Figure \ref{fig:BesselFit}(a) shows a spectrum of the interferometer output when the micromirror is driven with RF voltage amplitude of $v_{d}=10$V. The even order modulation sidebands are suppressed, and the relative peak heights of the odd order modulation sidebands can be determined. By fitting the peak heights of first order modulation sideband to $A\cdot J_1(s \cdot v_{d}^2)$ as a function of $v_{d}$, we obtained $s=0.025 \pm 0.0005 / V^2$.

We plot the peak heights of each odd order modulation sideband as a function of the RF voltage amplitude in Fig. \ref{fig:BesselFit}(b). The DC components corresponding to $J_0(s \cdot v_{amp}^2)$ are measured by scanning $\phi_i$ from 0 to $\pi/2$ for each RF amplitude, and fit using the $s$ value obtained from first order modulation sidebands. The amplitude of $J_0$ values have to be normalized since the DC and AC current are measured using different transimpedance gain in our circuit. The data fit well to Bessel functions with a single scaling parameter $s$ up to fifth order. A DC voltage of 135V is necessary to induce a phase shift of $\pi$ for the 780 nm laser for this mirror, but a smaller RF voltage amplitude of $v_{amp} \approx 11$ V is sufficient to generate phase oscillation between $\pm \pi$ when a mechanical resonance is utilized.

Phase modulation based on reflection off a micromirror works over a broad wavelength range, and is insensitive to polarization, power of the incoming light and potential thermal drifts from driving RF power. Using micromirrors with higher resonant frequencies, higher modulation frequencies can be achieved than by using macroscopic mirrors \cite{DebsAO2008}. A large array of such micromirror-based phase modulators can be fabricated in a single chip in the case where a scalable solution is desired. 

In this work, we showed that micromirrors can be used to construct beam steering devices with interferometric phase stability and ideal phase modulators. The authors would like to thank F. P. Lu at AQT for help with the micromirrors. This work was supported by ARO, office of the Director of National Intelligence, and Intelligence Advanced Research Projects Activity.

\end{document}